# Shrinkage anisotropy characteristics from soil structure and initial sample/layer size

V.Y. Chertkov*

Division of Environmental, Water, and Agricultural Engineering, Faculty of Civil and Environmental Engineering, Technion, Haifa 32000, Israel

**Abstract.** The objective of this work is a physical prediction of such soil shrinkage anisotropy characteristics as variation with drying of (i) different sample/layer sizes and (ii) the shrinkage geometry factor. With that, a new presentation of the shrinkage anisotropy concept is suggested through the sample/layer size ratios. The work objective is reached in two steps. First, the relations are derived between the indicated soil shrinkage anisotropy characteristics and three different shrinkage curves of a soil relating to: small samples (without cracking at shrinkage), sufficiently large samples (with internal cracking), and layers of similar thickness. Then, the results of a recent work with respect to the physical prediction of the three shrinkage curves are used. These results connect the shrinkage curves with the initial sample size/layer thickness as well as characteristics of soil texture and structure (both inter- and intra-aggregate) as physical parameters. The parameters determining the reference shrinkage curve (relating to the small samples) and initial sample size/layer thickness are needed for the prediction of the above soil shrinkage anisotropy characteristics. Using the available data on two soils and samples of two essentially different sizes, illustrative estimates of the relative sample sizes, new characteristic of the shrinkage anisotropy, and shrinkage geometry factor are given as the physically predicted values.

---

*Corresponding author. Tel.: 972-4829-2601.
*E-mail address:* agvictor@tx.technion.ac.il; vychert@ymail.com (V.Y. Chertkov).

**1. Introduction**

The anisotropy of soil shrinkage essentially influences soil hydraulic properties, water flow, and transport phenomena, even though there are no cracks (e.g., Garnier et al., 1997a, 1997b). It is obvious that cracking additionally complicates this influence (e.g., Coppola et al., 2012). For this reason the prediction of the soil shrinkage anisotropy characteristics based on the soil texture and structure (both inter- and intra-aggregate) as well as the initial sample/layer sizes as physical parameters, is important for the physical understanding hydrological processes in shrink-swell soils. Shrinkage anisotropy reflects the possible difference between the vertical and horizontal shrinkage of a sample. Starting from Bronswijk (1988, 1989, 1990, 1991a, 1991b), accounting for shrinkage anisotropy is usually conducted in terms of the shrinkage geometry factor, $r_s$. There are a number of works on the shrinkage geometry factor, its measurement, and use for different aims (Bronswijk, 1988, 1989, 1990, 1991a, 1991b; Garnier et al., 1997a, 1997b; Baer and Anderson, 1997; Crescimanno and Provenzano, 1999; Cornelis et al., 2006; Peng and Horn, 2007; Boivin, 2007; Coppola et al., 2012, among others). Also available are works on the necessary and essential corrections of the shrinkage geometry factor estimated by Bronswijk's approach (Chertkov, 2005, 2008a). However, $r_s$ remains the empirical engineering parameter that is a complex function of soil water content, $W$ (Chertkov, 2005, 2008a) and determined (at a given $W$) from measurements of soil subsidence and soil matrix shrinkage (or soil crack volume). Thus, it is not currently possible to physically predict this engineering parameter from soil structure, physical properties, and sample size or layer thickness. The objective of this work is to suggest such prediction of $r_s$ for soils with relatively small organic matter content (~0.1-3.5%) and, in addition, to introduce a more immediate presentation of soil shrinkage anisotropy through the sample/layer size ratios, also as a physically predicted function of $W$. First, we briefly review the results of a recent work (Chertkov, 2012a) with respect to the physical prediction of three different shrinkage curves of a soil, relating to small samples (without cracking at shrinkage), sufficiently large samples (with internal cracking), and layers of similar thickness (Section 2). These results are used in the following. We, then, consider the connection between the variation of sample/layer sizes with water content and three shrinkage curves (Section 3), new presentation of the soil shrinkage anisotropy concept (Section 4), physical prediction of $r_s$ and some of its applications (Section 5), some specifications connected with water content profile and horizontal cracks (Section 6). Finally, we give and discuss illustrative estimates for two soils and two initial sample sizes from Crescimanno and Provenzano (1999), with respect to the current relative sample/layer sizes, new characteristic of soil shrinkage anisotropy, and shrinkage geometry factor as physically predicted functions of water content (Section 7). Notation of the values that repeat is summarized at the end of the paper.

**2. Shrinkage curves as functions of soil structure and initial sample/layer size**

In a recent work Chertkov (2012a) considered an approach to the physical prediction of a soil shrinkage curve, including the crack volume contribution, depending on sample size or layer thickness as well as soil texture and structure for aggregated soils with negligible organic matter content. Three key points of the approach are: (i) accounting for the recently suggested intra-aggregate structure (Fig.1) including lacunar pores and an aggregate surface layer with specific properties (Chertkov, 2007a, 2007b, 2008b); (ii) the consideration of all the contributions to the soil volume and water content based on the inter- and intra-aggregate soil structure; and (iii) the use of new concepts of the *lacunar factor* ($k$), the *crack factor* ($q$), and *critical sample size* ($h^*$). The calculation of $k$ as a function of the ratio, $c/c_*$ ($c$ being the soil clay content; $c_*$ being the critical clay content (Chertkov, 2007a); in particular, $c_*$ depends on clay type) in the case where the cracks do not exist in small samples (Chertkov, 2010), was generalized to the case of large samples with internal cracking (Chertkov, 2012a). The crack factors, $q_s$ for samples and $q_l$ for layers as functions of the ratio, $h/h^*$ ($h$ being the initial sample size or layer thickness at maximum swelling) are calculated as (Chertkov, 2012a)

$q_s(h/h^*)=0,$   $0<h/h^*\leq 1$   (1a)

$q_s(h/h^*)=b_1(h/h^*-1)^2,$   $1\leq h/h^*\leq 1+\delta$   (1b)

$q_s(h/h^*)=1-b_2/(h/h^*-1),$   $h/h^*\geq 1+\delta$   (1c)

and

$q_l(h/h^*)=b_1(h/h^*)^2,$   $0\leq h/h^*\leq \delta$   (2a)

$q_l(h/h^*)=1-b_2/(h/h^*),$   $h/h^*\geq \delta$   (2b)

where

$$\delta=(1/(3b_1))^{1/2}, \qquad b_2=2\delta/3=(2/3)(1/(3b_1))^{1/2}. \qquad (3)$$

with theoretical values of universal constants, $b_1\cong 0.15$, $b_2\cong 1$, and $\delta\cong 1.5$. The critical sample size, $h^*$ is calculated as a function of minimum ($X_{min}$) and maximum ($X_m$) aggregate sizes, and structural porosity ($P_h$) at maximum swelling (Chertkov, 2012a). The shrinkage curves (specific volume vs. gravimetric water content) of the sample of a given size, $Y_s(W, h/h^*)$ and of the layer of a given thickness, $Y_l(W, h/h^*)$, including crack volume contribution, are expressed through the reference shrinkage curve, $Y_r(W)$ (Chertkov, 2007a, 2007b), and the $q$ factor as (Chertkov, 2012a)

$$Y_s(W, h/h^*)=(1-q_s(h/h^*))Y_r(W)+q_s(h/h^*)Y_{rh}, \qquad 0\leq W\leq W_h \qquad (4)$$

$$Y_l(W, h/h^*)=(1-q_l(h/h^*))Y_r(W)+q_l(h/h^*)Y_{rh}, \qquad 0\leq W\leq W_h \qquad (5)$$

where $Y_{rh}\equiv Y_r(W_h)$. The physical parameters determining the reference shrinkage curve, $Y_r(W)$ (including the $k$ factor) have been discussed in detail (Chertkov, 2007a, 2007b, 2010, 2012a).

## 3. Sample and layer size evolution at shrinkage

The found specific soil volumes, $Y_s$ and $Y_l$ (section 2) allow one to estimate the variation of the different sample and layer sizes at shrinkage. The layer sizes can be used to predict the subsidence of layer surfaces and horizontal deformation of the soil matrix inside the cracked layer. The sample sizes can also be used to consider shrinkage anisotropy at cracking (section 4).

First, we define and regard the current sizes of a sample. In the case of an initially cubic sample there are three sizes of interest (Fig.2): the current horizontal sample size, $x'(W)$ (with crack contribution at $h>h^*$ or without it at $h<h^*$); the current sample height, $z'(W)$; and the current horizontal size, $x''(W)$ of the matrix inside the sample if the summary crack volume is *mentally* excluded (at $h<h^*$ $x'=x''$). In the initial state $z'(W_h)=x'(W_h)=h$. Below we also use the layer volume in Bronswijk's approximation (Chertkov, 2005) when the layer at maximum swelling is constructed of a set of contacting, but disconnected cube samples. In such an approximation gaps appear between the initial cubes at the layer shrinkage process. Hence, the latter can be illustrated by the same Fig.2 for the separate cube sample. The specific soil volume of the layer in this approximation is designated as $Y_l'$ (unlike $Y_l$). We take advantage, below, of the link between $Y_l'$ and the specific soil volume, $Y_l$ of a real connected and cracked layer (Chertkov, 2005) as

$$Y_l=Y_l'(1+x'/h)^2/4. \qquad (6)$$

In this link (Eq.(6)) $x'$ is the current horizontal sample size (Fig.2a). In addition, the specific volume of the layer in Bronswijk's approximation, $Y_l'$ is simply connected to the current sample height, $z'$ (Fig.2c). Indeed, in such an approximation the current layer volume of the thickness $z'$ (per $h\times h$ surface area of the layer) can be written as $z'h^2$ (see Fig.2c) and by definition of $Y_l'$ as $Y_l'h^3/Y_h$ ($h^3/Y_h\equiv m$ is the oven-dried layer mass per $h\times h$ surface area of the layer). From the equality of these values one obtains

$$Y_l'=(z'/h)Y_h. \qquad (7)$$

Still another necessary relation between sample height, $z'$ and horizontal size, $x'$ (Fig.2),

$$(x'/h)^2(z'/h)=Y_s/Y_h \ , \qquad (8)$$

follows from the equality $x'^2z'=Y_s h^3/Y_h$ between the two different expressions of the current sample volume. Finally, we use the relation between the sample height, $z'$ and horizontal size $x''$ of the soil matrix inside the sample (after the *mental* exclusion of possible cracks) (Fig.2),

$$(x''/h)^2(z'/h)=Y_r/Y_h \ , \qquad (9)$$

that follows from the equality $x''^2z'=Y_r h^3/Y_h$ between the two different presentations of the current volume of the soil matrix inside the sample. The last four relations (Eqs.(6)-(9)) between the values that we seek: $z'$, $x'$, $x''$, and $Y_l'$, enable one to find those through $Y_s(W, h/h^*)$, $Y_l(W, h/h^*)$, $Y_r(W)$, $Y_h$, and $h$ as

$$z'/h=(Y_s/Y_h)[2(Y_l/Y_s)^{1/2}-1]^2 \ , \qquad (10)$$

$$x'/h=[2(Y_l/Y_s)^{1/2}-1]^{-1} \ , \qquad (11)$$

$$x''/h=(Y_r/Y_s)^{1/2}[2(Y_l/Y_s)^{1/2}-1]^{-1} \ , \qquad (12)$$

$$Y_l'/Y_h=(Y_s/Y_h)[2(Y_l/Y_s)^{1/2}-1]^2 \ . \qquad (13)$$

Note that at $Y_s=Y_r$ (when there are no cracks in the sample at $h<h_*$), $x'$ (Eq.(11)) coincides with $x''$ (Eq.(12)) as it should be. In addition, $z'/h=Y_l'/Y_h$ (Eqs.(10) and (13)), also as it should be.

Unlike the sizes of the sample ($z'$, $x'$, $x''$; Fig.2), two sizes characterizing layer shrinkage (Fig.3) are only expressed through $Y_l$, $Y_r$, and $Y_h$. These two sizes are (Fig.3) the current thickness of the real shrinking and cracking layer, $z$ and the current horizontal size $x$ of the soil matrix volume (per $h \times h$ surface area of the layer) after the *mental* exclusion of the crack volume. The expressions

$$z/h=Y_l/Y_h \qquad (14)$$

and

$$x/h=(Y_r/Y_l)^{1/2} \qquad (15)$$

follow, respectively, from the equality of the two different presentations of the current layer volume (per $h \times h$ surface area of the layer), $h^2z=Y_l h^3/Y_h$, and the equality of the two different presentations of the current volume of the soil matrix inside the layer (per $h \times h$ surface area of the layer) after the *mental* exclusion of crack volume, $x^2z=Y_r h^3/Y_h$.

Since the shrinkage curves, $Y_s(W,h/h^*)$ and $Y_l(W,h/h^*)$, depend on the initial sample/layer size, $h$ (see Eqs.(1)-(5)), the variation of $Y_l/Y_h$ and the relative sample/layer sizes, $z'/h$, $x'/h$, $x''/h$, $z/h$, and $x/h$ with the water content also depend on $h$ (see Eqs.(10)-(15)). Note that the above sample/layer sizes are physically predicted through the specific soil volumes (Eqs.(1)-(5)) *without* using the concept of the shrinkage geometry factor of the sample or layer (see section 5).

**4. The anisotropy of soil shrinkage**

The shrinkage geometry factor, $r_s$ is an indirect characteristic of soil shrinkage anisotropy. The rather more immediate and clear presentation and prediction of soil shrinkage anisotropy, $a(W, h/h^*)$ can be given by the three possible ratios of the vertical and horizontal sample and layer sizes as

$$a(W, h/h^*): z'/x', \quad z'/x'', \quad z/x \ . \tag{16}$$

The sizes are known from Eqs.(10) - (12), (14) and (15) as physically predicted functions of the water content, $W$. Figure 4 shows the possible qualitative view of the above ratios for a soil sample or layer. The view also depends on the $h/h^*$ ratio. In addition to the "external" shrinkage anisotropy, $z'/x'$, connected with the external sample sizes (Fig.2), the ratios $z'/x''$ and $z/x$ of sample ($z'$) and layer ($z$) vertical sizes to "internal" horizontal sizes of the soil matrix in sample ($x''$) and layer ($x$) (Figs.2 and 3), can be useful for the understanding and prediction of the anisotropy of cracking and the corresponding anisotropy of hydraulic conductivity in real soils.

Examples of the $a(W)$ function (i.e., the $z'/x'$, $z'/x''$, $z/x$ ratios) prediction, at the "small" ($h/h^*<1$) and "large" ($h/h^*>1$) $h$ values for real soils in sample and layer geometry, are considered in Section 7.

**5. The shrinkage geometry factor and some of its applications**

The corrected shrinkage geometry factor, $r_s$ (compared to Bronswijk's (1990) approximation) for the *sample* is written as (Chertkov, 2005, 2008a)

$$r_s(W, h/h^*)=\log(Y_r(W)/Y_h)/\log(z'(W, h/h^*)/h)=\log(Y_r(W)/Y_h)/\log(Y_l'(W, h/h^*)/Y_h) \ . \tag{17}$$

The physically predicted $Y_r(W)$, $Y_l'(W, h/h^*)$, and $z'(W, h/h^*)$ (sections 2 and 3) mean the similar prediction of $r_s$. Such a physically predicted $r_s$ value, for the sample case with possible cracks, can be used, for instance, in the problem of water flow in swelling soils in framework of Garnier et al.'s (1997a, 1997b) approach.

One should note the difference between the shrinkage of a small sample ($h/h^*<1$) and a large one with cracks ($h/h^*>1$) as applied to the $r_s$ concept. In the former case the decrease of the matrix volume of the soil sample consists of two contributions, sample volume decrease at the expense of vertical size decrease (subsidence) and at the expense of horizontal size decrease (lateral deformation). For this reason, knowing $r_s$ and, for instance, the vertical size, one can estimate the current horizontal size using Bronswijk's (1990) known formula. In the latter case the crack volume, occurring inside the sample, gives the additional third contribution (that is negative since crack volume grows at shrinkage) to the decrease of the matrix volume of the soil sample. Therefore, in the case of large samples ($h/h^*>1$) the knowledge of $r_s$ and the current vertical sample size are not sufficient for the separation between the contribution of the internal cracks and that of the lateral deformation. Note that the physical prediction of the sample $z'$, $x'$, and $x''$ sizes (Eqs.(10)-(12)) decides this issue.

The corrected shrinkage geometry factor, $r_s$ (compared to Bronswijk's (1990) approximation) relating to the *layer* case is expressed through the different specific volumes of the soil as (Chertkov, 2005, 2008a)

$$r_s(W, h/h^*)=\log(Y_r(W)/Y_h)/\log(Y_l(W, h/h^*)/Y_h) \tag{18}$$

($Y_h$ is the specific soil volume at maximum swelling). All values entering this expression are already known (see above). Since the values entering Eq.(18) depend on a number of physical soil parameters (Chertkov, 2012a), this equation physically predicts the $r_s(W)$ dependence for a *cracked soil layer* (with water content being homogeneous within the layer).

Note that unlike the sample case (see above), in a layer of any thickness there are only two contributions to the decrease in the matrix volume of the soil layer at shrinkage, layer thickness decrease (a positive contribution) leading to soil subsidence and crack volume increase in the layer (a negative contribution). Therefore, in the layer case (field conditions), knowing $r_s$ (e.g., from Eq.(18)) and the soil subsidence along the vertical profile, one can predict the crack volume in the layers using Bronswijk's (1990) formula. However, the crack volume can be physically predicted without the use of the $r_s$ concept (Chertkov, 2012a). In light of that, it is worth noting another *possible and important application* of the physical prediction of the $r_s$ value (Eq.(18)) to estimating some aspects of crack network geometry in a layer. Indeed, the physical prediction of the shrinkage geometry factor as a function of $W$ in the layer case enables a similar prediction of the crack-width distribution in the layer (Chertkov, 2008a) The crack-width distribution is important for physical estimating the soil hydraulic properties.

Examples of the $r_s(W)$ dependence prediction at the small and large $h/h^*$ values for real soils in sample and layer geometry are considered in Section 7.

**6. The effects of water content profile variation and horizontal cracks**

We implied above that water content is homogeneous within the limits of a layer and sample. For this reason, when estimating the crack network characteristics in the vertical cross-section of a soil, one should divide the cross-section into horizontal layers with water content being homogeneous within the limits of each layer. Such a division is carried out using the observed (or predicted) water content profile. However, this profile can vary with time. As a result, a layer with initially homogeneous water content should be divided after some time into two thinner layers or, on the contrary, this layer can enter a thicker layer. Therefore, variations in water content profile can lead to such observable effects as, e.g., the increase of the total crack volume in a layer at drying with the occurrence (or opening) of larger cracks and the closing of many small cracks that appeared earlier (Hallaire, 1984). In any case for the dependable prediction of the evolution in thickness of soil layers, and crack volume or geometry (i.e., different crack distributions) inside them, one needs to use the sufficiently accurate data on the vertical water content profile (or prediction of the profile) and its variation with time.

Vertical cracks were considered above. Horizontal cracks are secondary ones since they start from the walls of the vertical cracks as a result of additional drying and shrinkage of the soil matrix along the walls (Chertkov and Ravina, 1999). The horizontal cracks contribute little to the total crack volume (Chertkov, 2008a). By definition of the reference shrinkage curve, $Y_r(W)$ the horizontal cracks do not influence the latter. In addition, the development of horizontal cracks neither changes the soil ($Y_l$) and vertical crack ($U_{cr\ l}$) volumes, nor the soil surface subsidence (Chertkov and Ravina, 1999; Chertkov, 2008a). For this reason the horizontal cracks have little influence on the shrinkage geometry factor, $r_s$ (Eq.(18)) and distributions of the vertical crack characteristics. Note, however, that contribution of horizontal cracks to the soil hydraulic conductivity (that is beyond the scope of this work) can be essential.

Finally, it should be emphasized that the above noted vertical and horizontal cracks are macroscopic those (up to tens of centimeters in size) and form the shrinkage crack network (for the 2D illustrative example see Guidi et al., 1978) unlike the microcracks that are observed on the images of the thin soil sections and can have any orientation (e.g., Bui and Mermut, 1989; Velde et al., 1996). Merging of the microcracks under shrinkage stresses leads to the network formation of the quasi vertical and quasi horizontal macrocracks through the mechanism of multiple cracking (Chertkov and Ravina, 1998; Chertkov, 2008a).

**7. Illustrative estimates of the relative sizes, shrinkage anisotropy, and shrinkage geometry factor of samples and layers**

The aim of this section is to use two soils from Crescimanno and Provenzano (1999) that were considered in Chertkov (2012a) as examples, in order to illustrate the possible behavior of the soil shrinkage anisotropy characteristics, regarded above for sample and layer geometry, as functions of water content and initial sample/layer size. The illustrative dependences in Figs.5-7 rely on the shrinkage curves, $Y_r(W)$ (Chertkov, 2007a, 2007b), $Y_s(W, h/h^*)$ and $Y_l(W, h/h^*)$ (Eqs.(1)-(5)) (Chertkov, 2012a) that were found for Delia 1a and Delia 6 soils. Delia 2a and Delia 4 soils from Crescimanno and Provenzano (1999) that were also considered in Chertkov (2012a), lead to similar dependencies. The soil physical characteristics that are necessary to predict $Y_r(W)$, $Y_s(W, h/h^*)$, and $Y_l(W, h/h^*)$ for the two soils are indicated in Table 1.

The curves in Figs.5-7 relating to samples and layers are marked by solid and dashed lines, respectively. For each shrinkage characteristic (except for $x''/h$ and $z'/x''$; see below) two curves are shown that correspond to small ($h$=3.1 cm $<h^*$) and large ($h$=11.5 cm $>h^*$) sample or layer size (for $h^*$ see Table 1). These sizes coincide with those from Crescimanno and Provenzano (1999).

The *relative sample sizes* from Eqs.(10)-(12) (see also Fig.2) and *relative layer sizes* from Eqs.(14) and (15) (see also Fig.3) predicted for Delia 1a and Delia 6 soils are shown in Fig.5. At $h$=3.1 cm $<h^*$ $x''=x'$. For this reason $x''/h$ is only shown at $h$=11.5 cm $>h^*$. Scrutinizing Fig.5 one notes the following points.

1. The different characteristic sizes of the sample or layer can coincide with each other at some water content values (except for the point $W=W_h$).

2. The relative characteristic sizes of small ($h<h^*$) samples (curves 4 and 5) and thin ($h<h^*$) layers (curves 8 and 9) meet the inequality

$$z/h < z'/h < x'/h < x/h \quad . \tag{19}$$

That is, in the case of small samples and layers: (i) the vertical shrinkage (relative subsidence 1-$z'/h$ or 1-$z/h$) dominates the lateral (1-$x'/h$ or 1-$x/h$) one (which corresponds to cracking in the layer case); and (ii) this domination is stronger for layers than for samples:

$$x/h - z/h > x'/h - z'/h \quad . \tag{20}$$

This is natural since the internal tension or stretching occurs in a shrinking layer (Chertkov, 2005). The values of $x'/h$, $z'/h$, $x/h$, and $z/h$ (or positions of curves 4, 5, 8, and 9 in Fig.5) themselves depend on the specific features of the particular soil (see Table 1).

3. In the case of large samples and layers ($h>h^*$) there are two general laws: (i) curve 1 is higher than curve 2, i.e., $x'/h > x''/h$, this is just the consequence of the $x'$ and $x''$ definitions in section 3 (Fig.2); and (ii) curves 3 and 7 lie higher than curves 5 and 9, i.e., the relative subsidence, 1-$z'/h$ or 1-$z/h$ of small samples (curve 5) or thin layers (curve 9) exceeds that of large samples (curve 3) or thick layers (curve 7). In all other relations the mutual arrangement of curves 1-3, 6, and 7 of large samples and layers ($h>h^*$) and values of their relative characteristic sizes are determined by the specific features of the particular soil (see Table 1) and the relative initial size ($h/h^*$).

The *shrinkage anisotropy* characteristics, $z'/x'$, $z'/x''$, and $z/x$, predicted for Delia 1a and Delia 6 soils are shown in Fig.6. $z'/x''$ is only shown at $h$=11.5cm $>h^*$ as $z'/x''=z'/x'$ at $h$=3.1cm $<h_*$. The following points about the shrinkage anisotropy curves in Fig.6 are worth noting.

1. Different shrinkage anisotropy curves can intersect (this is the consequence of the similar property of the curves in Fig.5).
2. Deflection of curves in Fig.6 with respect to the isotropy case ($z'/x'=z'/x''=z/x=1$) is maximum for thin layers (curve 5) and then, for small samples (curve 3). In addition, this deflection for small samples and thin layers increases with drying. At qualitative similarity between the anisotropy curves of small samples and thin layers (curves 3 and 5, respectively) for the different soils, the curves quantitatively and appreciably differ from each other depending on the particular soil features (see Table 1).
3. Unlike in the case of small samples and thin layers, the behavior of shrinkage anisotropy curves for large samples (curves 1 and 2) and thick layers (curve 4) qualitatively varies from soil to soil. Indeed, both $z'(W)<x'(W)$ and $z'(W)>x'(W)$ are possible (cf. curves 1 for Delia 1a and Delia 6 soils) and both $z(W)<x(W)$ and $z(W)>x(W)$ are possible (cf. curves 4 for Delia 1a and Delia 6 soils). At the same time, for both soils $z'(W)>x''(W)$ (see curves 2 for Delia 1a and Delia 6 soils). In addition, for both soils curve 2 of a large sample is higher than another curve of a large sample (curve 1) since, by definition, $x'>x''$. The similar mutual arrangement of curves 2 (for a large sample) and 4 (for a thick layer), i.e., the inequality, $z'(W)/x''(W)>z(W)/x(W)$ is also observed for the two soils, but this result is not some trivial consequence of a definition.

The *shrinkage geometry factors* for samples (Eq.(17)) and for layers (Eq.(18)) predicted for Delia 1a and Delia 6 soils, are shown in Fig.7. A number of points should be noted in connection with Fig.7.

1. The mutual arrangement of the shrinkage-geometry-factor curves for small (curve 1) and large (curve 2) samples as well as thin (curve 3) and thick (curve 4) layers, is similar for the two soils.
2. Judging by curves 1 and 3 the volume shrinkage of small samples (curve 1) and thin layers (curve 3) principally occurs at the expense of the soil subsidence, since all curves 1 and 3 in Fig.7 correspond to $r_s$ values close to unity. Note that such predicted behavior is natural and expectable. With that the relative subsidence of a thin layer is larger than that of a small sample (for both soils) because curves 1 lie higher than curves 3 in Fig.7. This prediction is also reasonable.
3. For each of the two soils the appreciable fraction (~50%) of volume shrinkage of a thick-layer matrix or large-sample matrix turns into the crack volume since curves 2 and 4 in Fig.7 correspond to $r_s \cong 4 - 5$.
4. The specific course of each $r_s(W)$ dependence in Fig.7 is determined by the particular physical features of a corresponding soil (Table 1), sample/layer geometry, and relative size, $h/h^*$.

In general, one can state that the above considered soil shrinkage anisotropy characteristics, as functions of water content, can and should be predicted from the particular physical soil features of (i) the texture and intra-aggregate structure, (ii) inter-aggregate structure, and (iii) initial sample size or layer thickness (see Table 1 and Chertkov (2012a)).

## 8. Results and conclusion

In this work we considered the physical prediction of soil shrinkage anisotropy characteristics based on soil texture and structure (both inter- and intra-aggregate) as well as the initial sample/layer size. The results are as follows.

(i). *Sample and layer size evolution at shrinkage.* The shrinkage curves of sample, $Y_s(W, h/h^*)$ and layer, $Y_l(W, h/h^*)$ (Chertkov, 2012a) together with the reference shrinkage curve $Y_r(W)$ (Chertkov, 2007a, 2007b) allowed us to find the evolution of a number of characteristic sizes of soil sample and layer at shrinkage (vertical and horizontal size, and size of soil matrix without cracks). Figure 5 and the points noted in Section 7 illustrate the size dependencies on water content for two particular soils.

(ii). *Direct presentation of the shrinkage anisotropy through the sample size ratios.* The predicted evolution of vertical and horizontal sizes permitted us to suggest a new, more direct and visual presentation of the shrinkage anisotropy concept as the ratios of the sizes. The illustrative numerical estimates of shrinkage anisotropy (Fig.6) and the points noted in Section 7 show the potential usefulness and informativeness of the presentation.

(iii). *Presentation of the shrinkage geometry factor of samples and layers through soil structure and initial sample/layer size.* Knowing the $Y_s(W, h/h^*)$ and $Y_l(W, h/h^*)$ curves, one can predict the shrinkage geometry factor $r_s$ to be a function of physical soil characteristics. The illustrative numerical estimates of the shrinkage geometry factor for two particular soils (Fig.7) and the points noted in Section 7, show the evidence in favor of that. In turn, the predicted $r_s$ value determines the physically realizing combination of the crack volume and subsidence (as functions of water content) for a given soil layer and local hydrological conditions. According to Chertkov (2008a) the shrinkage geometry factor and crack network geometry (i.e., different distributions) in a layer are closely connected at shrinkage. This means that the crack network geometry can also be totally physically predicted using the soil structure and initial layer thickness.

The obtained results enable the physical prediction of all of the shrinkage anisotropy characteristics based on indicated soil and sample features (see Table 1). A recent physical prediction of the soil swelling curve (Chertkov, 2012b) suggests future extension of the above results to the cases of swelling and shrink-swell cycling.

**Notation**

- $a$     shrinkage anisotropy (Eq.(16)) (dimensionless)
- $b_1, b_2$   universal constants in Eq.(3) (dimensionless)
- $c$     clay content (dimensionless)
- $c_*$    critical clay content (dimensionless)
- $h$     initial sample size of approximately cubic shape at maximum swelling (cm)
- $h^*$   critical sample size at maximum swelling (cm)
- $k$     lacunar factor (dimensionless)
- $m$   oven-dried layer mass per $h \times h$ surface area of the layer (kg)
- $q$     crack factor (dimensionless)
- $q_l$    crack factor of the layer of initial thickness $h$ (dimensionless)
- $q_s$   crack factor of the approximately cubic- or cylindrically-shaped sample (with close height and diameter) (dimensionless)
- $r_s$   shrinkage geometry factor (dimensionless)
- $W$   gravimetric water content of soil (kg kg$^{-1}$)

| | |
|---|---|
| $W_h$ | $W$ value at maximum swelling (kg kg$^{-1}$) |
| $x$ | current horizontal size of the soil matrix at layer shrinkage inside the $h \times h$ basis (Fig.3) (cm) |
| $x'(W)$ | current horizontal size of the initial cubic sample (cm) |
| $x''(W)$ | size of soil matrix inside the sample (see Figs.2b and 2c) (cm) |
| $Y$ | specific volume of the soil with cracks (dm$^3$ kg$^{-1}$) |
| $Y_h$ | specific soil volume at maximum swelling (dm$^3$ kg$^{-1}$) |
| $Y_l$ | specific soil volume in the case of cracked, but connected layer (dm$^3$ kg$^{-1}$) |
| $Y_r(W)$ | reference shrinkage curve (dm$^3$ kg$^{-1}$) |
| $Y_{rz}$ | minimum specific volume of the reference shrinkage curve (dm$^3$ kg$^{-1}$) |
| $Y_s$ | specific soil volume in the case of a sample with or without cracks (dm$^3$ kg$^{-1}$) |
| $Y_l'$ | specific soil volume of the layer in Bronswijk's approximation (dm$^3$ kg$^{-1}$) |
| $z(W)$ | current thickness of the real shrinking and cracking layer (Fig.3) (cm) |
| $z'(W)$ | current height of the initially (approximately) cubic sample (Fig.2) (cm) |
| $\delta$ | universal constant in Eq.(3) (dimensionless) |

**Figure captions**

**Fig.1.** The schematic illustration of the accepted soil structure (Chertkov, 2007a, 2007b). Shown are (1) an assembly of many soil aggregates and inter-aggregate pores contributing to the specific soil volume, $Y$; (2) an aggregate, as a whole, contributing to the specific volume $U_a=U_i+U'$; (3) an aggregate indicated with two parts: (3a) interface layer contributing to the specific volume $U_i$ and (3b) intra-aggregate matrix contributing to the specific volumes $U$ and $U'=U/K$; (4) an aggregate indicated with the intra-aggregate structure: (4a) clay, (4b) silt and sand grains, and (4c) lacunar pores; and (5) an inter-aggregate pore leading, at shrinkage, to inter-aggregate crack contributing to the specific volume $U_{cr}$. $U$ is the specific volume of the intra-aggregate matrix (per unit mass of the oven-dried matrix itself). $U'$ is the specific volume of the intra-aggregate matrix (per unit mass of the oven-dried soil). $U_i$ is the specific volume of the interface layer (per unit mass of the oven-dried soil). $U_{cr}$ is the specific volume of cracks (per unit mass of the oven-dried soil). $U_a$ is the specific volume of aggregates (per unit mass of the oven-dried soil). $K$ is the aggregate/inter-aggregate mass ratio.

**Fig.2.** The cubic sample shrinkage of initial size $h$ at maximum swelling. **(a)** Horizontal cross-section. $x'$ is the current horizontal sample size at $W<W_h$. Possible cracks (black strips) are distributed in the shrinking sample. **(b)** The same horizontal cross-section. $x''$ is the current horizontal size of the matrix inside the sample at $W<W_h$. Shaded bands correspond to the summary crack volume (cracks are *mentally* collected together compared to Fig.2a). **(c)** The vertical cross-section of the shrinking sample. $z'$ is the current vertical sample size at $W<W_h$. The vertical shaded band corresponds to the summary crack volume (cracks are *mentally* collected together).

**Fig.3.** The layer shrinkage of initial $h$ thickness at maximum swelling. **(a)** The horizontal cross-section of the $h\times h$ basis at $W<W_h$. Black strips symbolize cracks. **(b)** The same horizontal cross-section of the $h\times h$ basis at $W<W_h$. $x$ is the current horizontal size of the soil matrix inside the $h\times h$ basis at $W<W_h$. Shaded bands correspond to the summary crack volume (cracks are *mentally* collected together compared to Fig.3a). $x>x''$ in Fig.2. **(c)** The vertical cross-section of the shrinking layer within the limits of the $h\times h$ basis (overlapping Fig.2c for comparison; $x'$, $z'$ and $x''$ are as in Fig.2c). $x$ is as in Fig.3b. $z$ is the current thickness of the real shrinking and cracking layer at $W<W_h$. $z<z'$ in Fig.2.

**Fig.4.** Possible qualitative view of the $a(W)$ dependence ($z'(W)/x'(W)$ or $z'(W)/x''(W)$ or $z(W)/x(W)$) of a soil sample or layer. Solid line: $a>1$. Dashed line: $a\leq 1$.

**Fig.5.** The relative sample/layer sizes of the two soils. The solid lines correspond to *sample* case: 1- $x'(W)/h$ dependence, $h=11.5$ cm$>h^*$; 2 - $x''(W)/h$, $h=11.5$ cm; 3 - $z'(W)/h$, $h=11.5$ cm; 4 - $x'(W)/h$, $h=3.1$ cm$<h^*$; 5 - $z'(W)/h$, $h=3.1$ cm. The dashed lines correspond to *layer* case: 6 - $x(W)/h$ dependence, $h=11.5$ cm; 7 - $z(W)/h$, $h=11.5$ cm; 8 - $x(W)/h$, $h=3.1$ cm; 9 - $z(W)/h$, $h=3.1$ cm.

**Fig.6.** The shrinkage anisotropy of the two soils. The solid lines correspond to *sample* case: 1- $z'(W)/x'(W)$ dependence, $h=11.5$ cm$>h^*$; 2 - $z'(W)/x''(W)$, $h=11.5$ cm; 3 - $z'(W)/x'(W)$, $h=3.1$ cm$<h^*$. The dashed lines correspond to *layer* case: 4 - $z(W)/x(W)$ dependence, $h=11.5$ cm; 5 - $z(W)/x(W)$, $h=3.1$ cm.

**Fig.7.** The shrinkage geometry factor of the two soils. The solid lines correspond to *sample* case: 1- $h=3.1$ cm$<h^*$; 2 - $h=11.5$ cm$>h^*$. The dashed lines correspond to *layer* case: 3 - $h=3.1$ cm; 4 - $h=11.5$ cm.

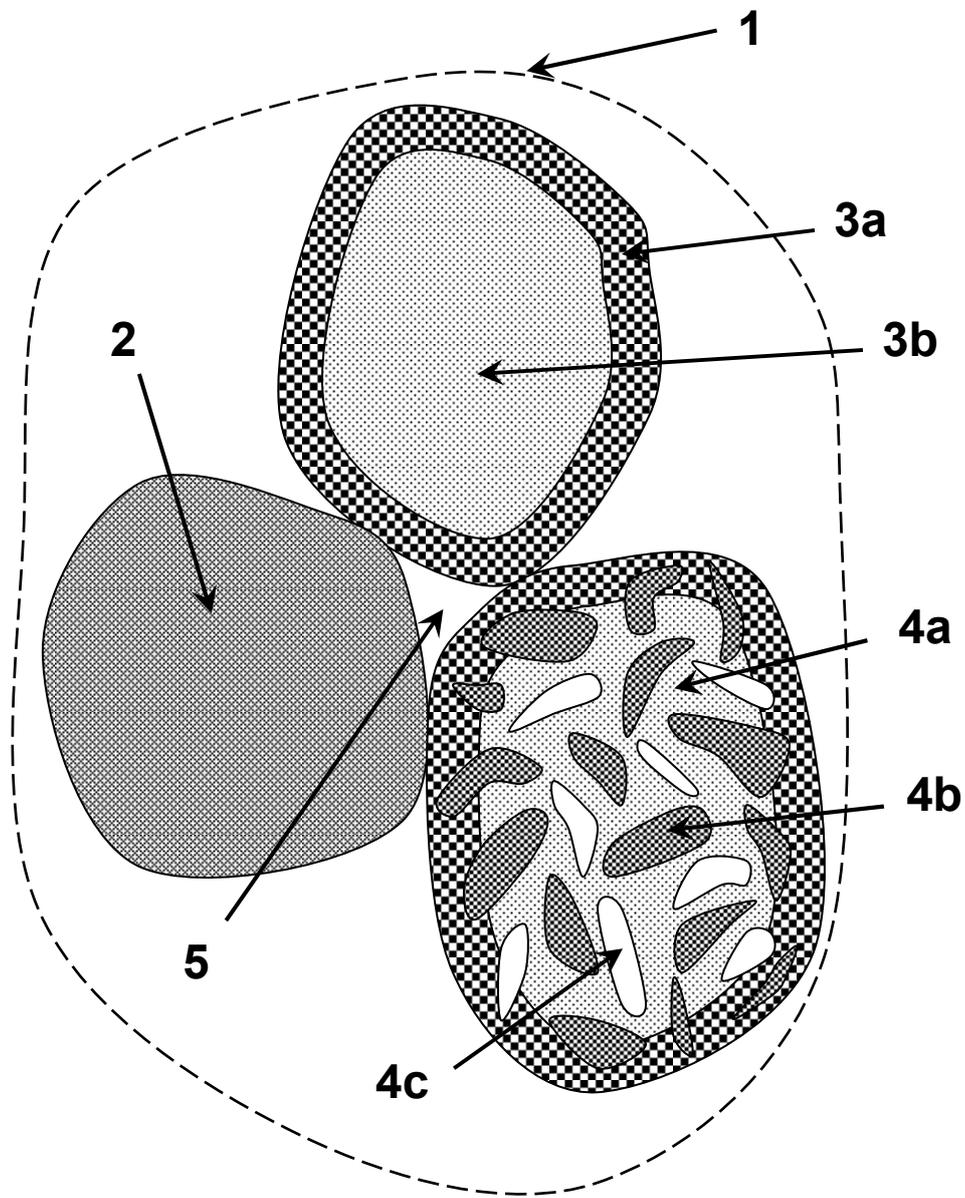

Fig.1

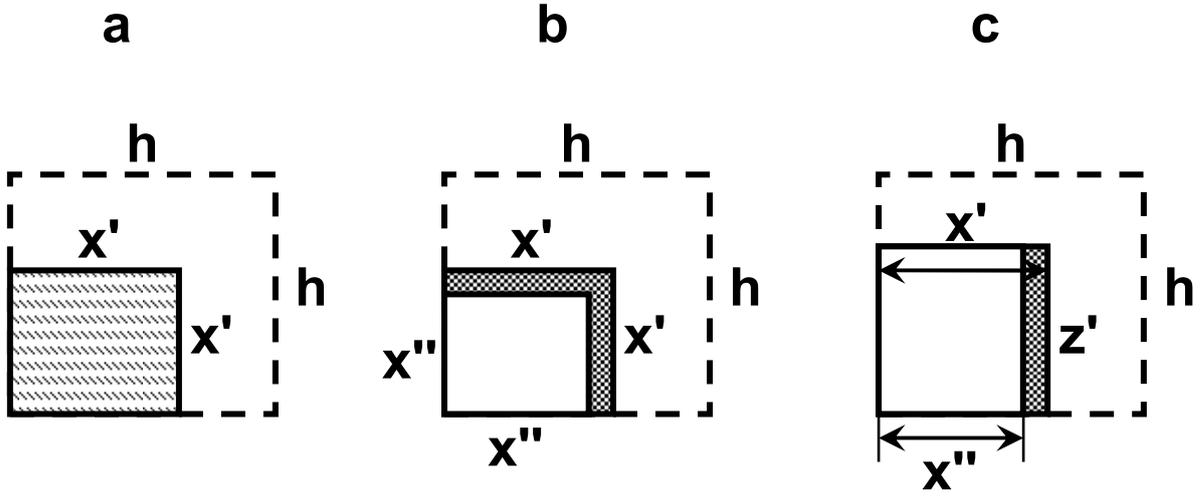

Fig.2

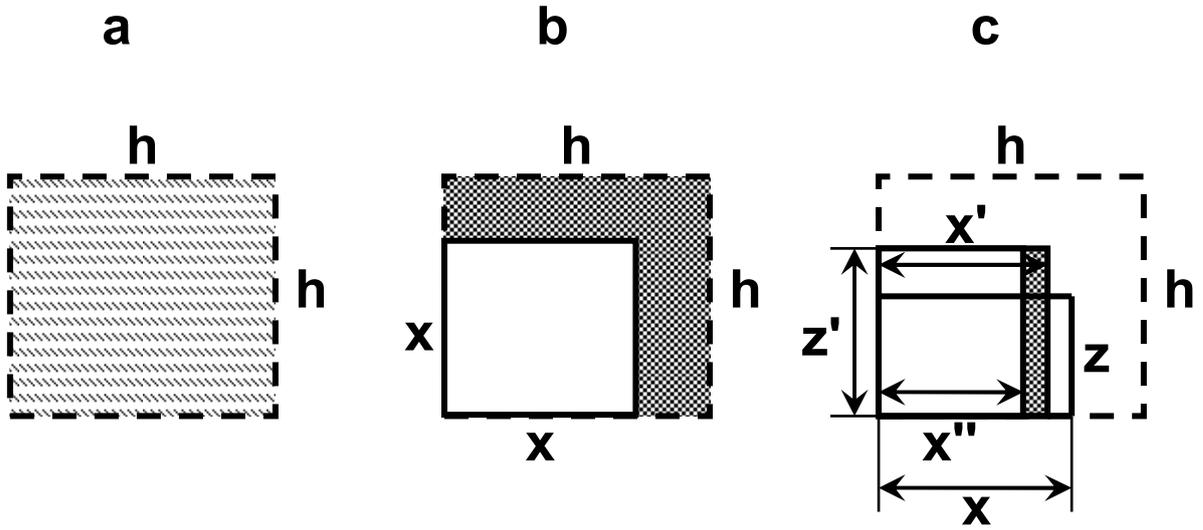

Fig.3

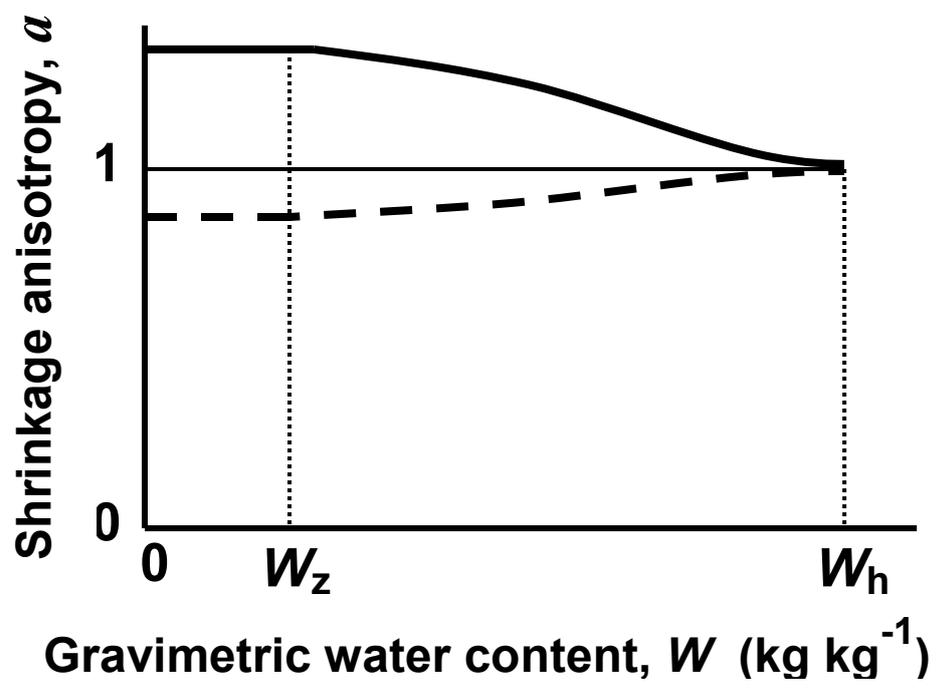

Fig.4

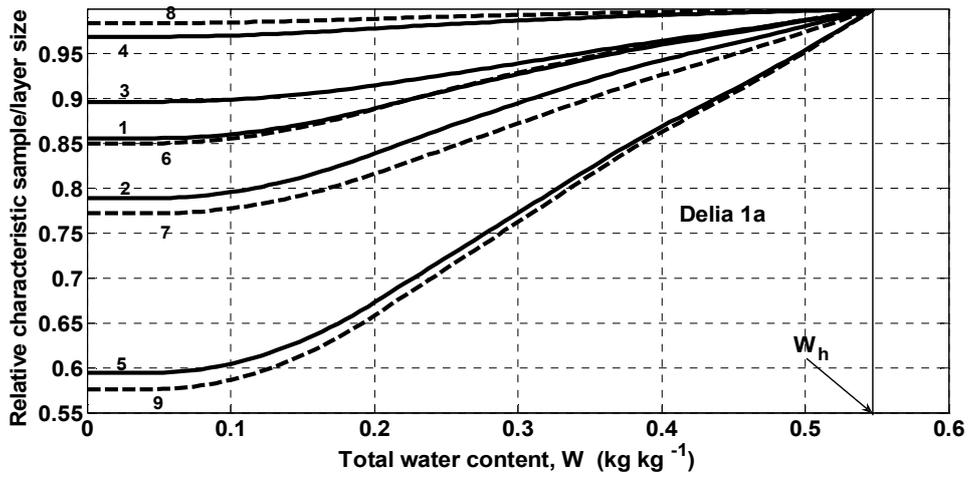
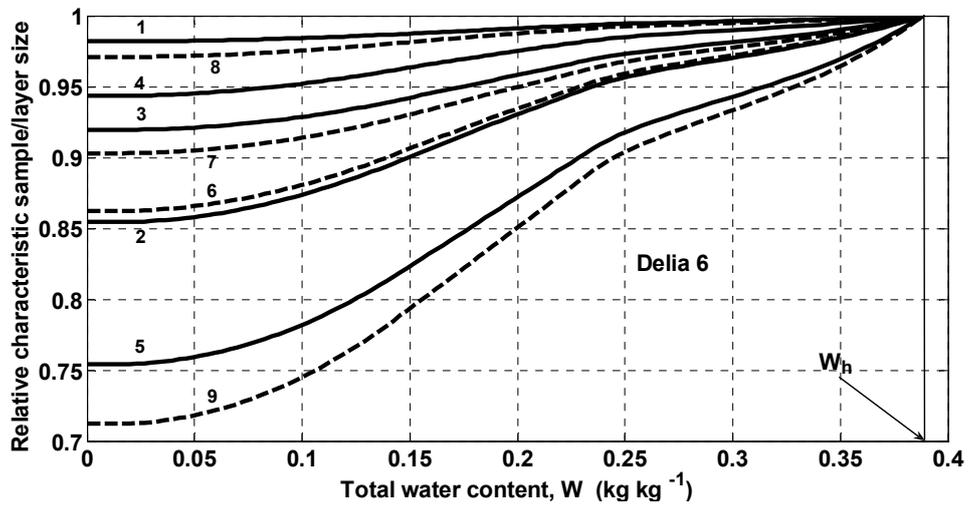

Fig.5

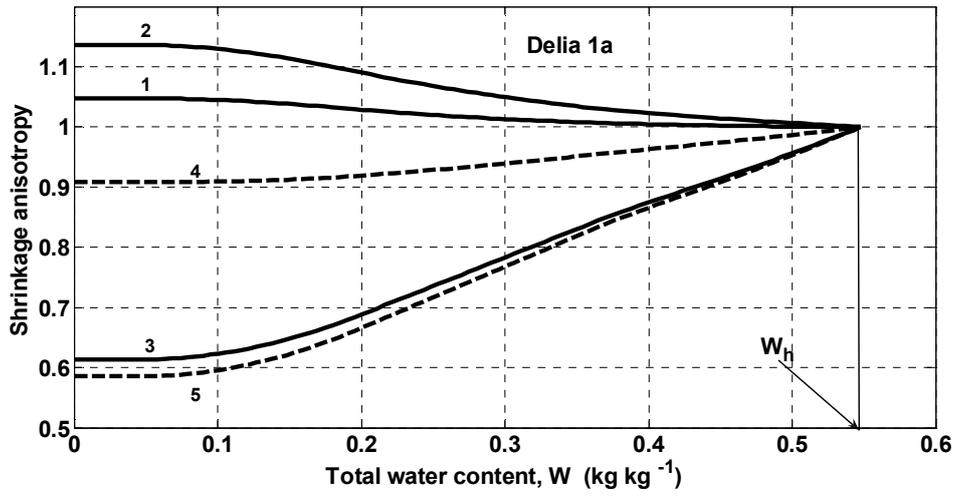

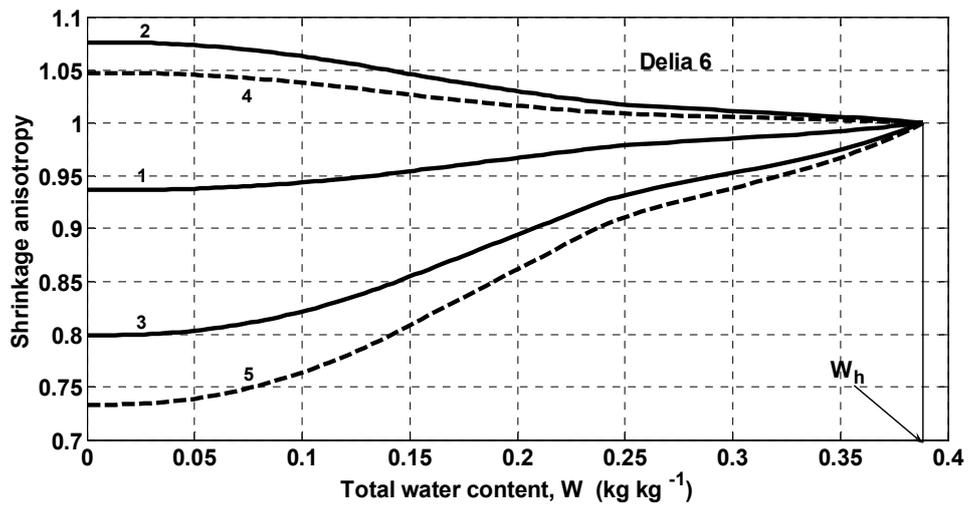

Fig.6

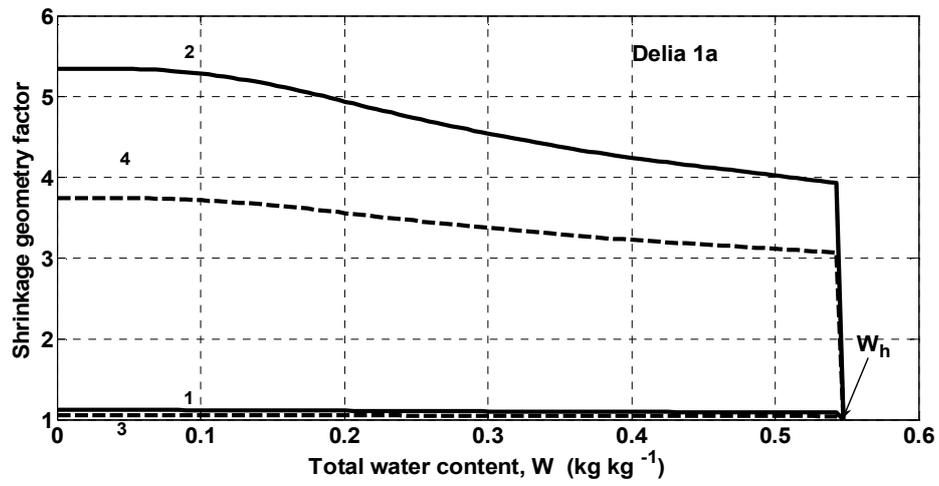

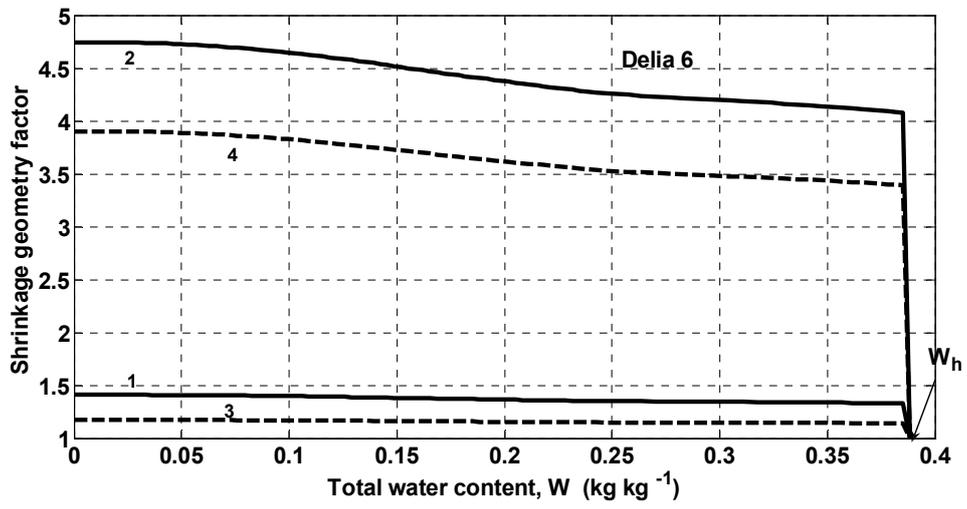

Fig.7

**Table 1.** Input parameters* for physical prediction of the reference shrinkage curve, $Y_r(W)$, sample shrinkage curve, $Y_s(W, h/h^*)$, and layer shrinkage curve, $Y_l(W, h/h^*)$ for two soils. The clay content values are from Crescimanno and Provenzano (1999); the other parameters were estimated in Chertkov (2012a) based on different data from Crescimanno and Provenzano (1999)

| Data source | $c$ | $\rho_s$ (kg dm$^{-3}$) | $Y_{rz}$ (dm$^3$kg$^{-1}$) | $W_h$ (kg kg$^{-1}$) | $k$ | $P_z$ | $U_{lph}$ (dm$^3$kg$^{-1}$) | $X_{min}$ (mm) | $X_m$ (mm) | $h^*$ (cm) |
|---|---|---|---|---|---|---|---|---|---|---|
| Crescimanno and Provenzano (1999, Table 1, Fig.4), soil Delia 1a | 0.57 | 2.6673 | 0.5149 | 0.5476 | 0 | 0 | 0 | 0.015 | 5.51 | 6.0 |
| As above, soil Delia 6 | 0.18 | 2.6428 | 0.5150 | 0.3886 | 0 | 0 | 0 | 0.052 | 4.21 | 3.4 |

* Clay content ($c$), mean solid density ($\rho_s$), minimum specific volume of the reference shrinkage curve ($Y_{rz}$), water content at maximum swelling ($W_h$), lacunar factor ($k$), structural porosity in oven-dried state ($P_z$), specific lacunar pore volume at maximum swelling ($U_{lph}$), minimum aggregate size ($X_{min}$), maximum aggregate size at maximum swelling ($X_m$), critical sample size ($h^*$).